\documentstyle[preprint,aps]{revtex}

\begin{document}
\draft
\preprint{}

\title{Premelting of Thin Wires}
\author{O. G{\"u}lseren, F. Ercolessi and E. Tosatti}
\address{
Scuola Internazionale Superiore di Studi Avanzati (SISSA), \\
Via Beirut 4, I-34014 Trieste, Italy
}
\date{\today}
\maketitle

\begin{abstract}
Recent work has raised
considerable interest on the nature of thin metallic wires.
We have investigated the melting behavior of thin cylindrical Pb wires
with the axis along a (110) direction,
using molecular dynamics and a well-tested many-body potential.
We find that---in analogy with cluster melting---the melting
temperature $T_m (R)$ of a wire with radius $R$ is lower than
that  of a bulk solid, $T_m^b$, by
$T_m (R) = T_m^b -c/R$.
Surface melting effects, with formation of a thin skin of highly
diffusive atoms at the wire surface, is
observed. The diffusivity is lower where the wire surface
has a flat, local (111) orientation, and higher at
(110) and (100) rounded areas.
The possible relevance to recent results on non-rupturizing thin necks
between an STM tip and a warm surface is addressed.

\end{abstract}
\pacs{PACS numbers:64.70.Dv,82.65.Dp,61.46.+w }

It is well known---experimentally and theoretically---that small
atomic clusters have a melting temperature $T_m$ which is lower
than that, $T_m^b$, of the infinite bulk solid of the same
element \cite{borel}. The main physical reason for this is that
the liquid-vapor interface free energy $\gamma_{lv}$ is generally
lower than the average solid-vapor interface free energy $\gamma_{sv}$.
Therefore, when the particle size is decreased and its surface to
volume ratio increases, the ``bulk'' liquid free energy at $T_m$ must
become higher than the ``bulk'' solid free energy to compensate for
the advantage of the liquid at the surface, and $T_m$ must decrease.

Thin wires with a microscopic diameter should exhibit a similar
melting behavior---although with significant differences due to
their cylindrical symmetry and infinite extension along the
axis---but they have not been extensively investigated so far,
probably due to lack of practical motivations. However, considerable
interest in the thermal behavior of thin wires has been recently raised
by some Scanning Tunneling Microscopy (STM) \cite{stm} experiments.
Abrupt changes of the tunnel resistance have been observed when the
tip--surface separation distance becomes sufficiently small, indicating
jump into contact \cite{durig}. Molecular dynamics (MD) computer simulation
studies \cite{uzi1,bell,tomagnini} have indicated that contact is often
realized by means of a narrow ``neck'' connecting the tip with the sample
surface.
In an MD study of a Au tip interacting with a Pb(110) surface at
high temperature\cite{tomagnini}, this neck is seen to be liquid in nature
in its thinner portion, even when the temperature is considerably lower
than $T_m$. Moreover, the jump to contact between tip and surface via such
a partly liquid neck was predicted, because of this, to occur more readily
as temperature was raised.
Recent STM experiments by Kuipers {\em et al.} \cite{kuipers,kuipers1},
also performed on Pb(110) using a W tip over a temperature range from
about $0.5\,T_M$ to surface melting, indicate that the neck elongates
up to considerable lengths (several thousands of $\rm\AA$), depending
on the retraction speed and on temperature, and then it breaks.
To shed some light on the behavior of such a thin and long neck, we have
started an investigation of cylindrical metallic wires.

{\bf (i) Simple theory:} For spherical particles of radius $R$, a melting
temperature $T_m(R)$ is obtained phenomenologically \cite{pawlow} by
equating the Gibbs free energies of solid and liquid spherical clusters,
assuming constant pressure conditions:
\begin{equation}
\label{sctm}
\frac{T_m^b - T_m(R)}{T_m^b} =
\frac{2}{\rho_s L^b R} \left( \gamma_{sv} -
\left( \frac{\rho_s}{\rho_l} \right)^{2/3}
\gamma_{lv} \right)
\end{equation}
where $\rho_s$ and $\rho_l$ are the solid and liquid densities,
$L^b$ is the bulk latent heat of melting and $\gamma_{sv}$ and
$\gamma_{lv}$ are the solid--vapor and liquid--vapor interface
energies respectively.
The surface energy anisotropy of the solid is not taken into account,
and the possibility of inhomogeneous phases (such as a liquid skin wetting
the solid cluster) is also neglected.
In spite of these approximations, the $1/R$ behavior is approximately
correct for clusters of sufficiently large size, although in some systems
$T_m$ drops faster than predicted \cite{borel,buffat,andreoni}.
In the case of Pb, which is the subject of our MD study, deviations from
$\sim 1/R$ appear to be small even at small sizes \cite{french,lim},
For a wire, following a similar procedure, we equate the Gibbs free
energies per unit length of the solid and liquid wires at constant
pressure and temperature, $G_s = N \mu_s + 2\pi R \gamma_{sv}$
and $G_l = N \mu_l + 2\pi R \gamma_{lv}$, where $N$ is the number of atoms
per unit length, $\mu_s$ and $\mu_l$ are the chemical potentials of the
solid and liquid phases respectively, and $R$ is the wire radius. Since
near $T_m^b$, $N ( \mu_s - \mu_l ) = V \rho_s L^b (T-T_m^b)/T_m^b$,
we obtain for the melting temperature $T_m(R)$ of a thin wire
\begin{equation}
\label{cctm}
\frac{T_m^b - T_m(R)}{T_m^b} =
\frac{1}{\rho_s L^b R} \left( \gamma_{sv} -
\left( \frac{\rho_s}{\rho_l} \right)^{1/2}
\gamma_{lv} \right)
\end{equation}
Hence, the melting temperature of a wire should be depressed,
approximately, by half the corresponding depression of a spherical
cluster. In this simple model the latent heat of melting per atom
(averaged on all atoms) decreases with exactly the same law when the
size is decreased:
\begin{equation}
\label{ccl}
\frac{L^b-L(R)}{L^b} = \frac{1}{\rho_s L^b R}
\left( \gamma_{sv} -
\left( \frac{\rho_s}{\rho_l} \right)^{1/2}
\gamma_{lv} \right)
\end{equation}
These relations imply the existence of a critical radius
$R_c = ( \gamma_{sv} - ( \frac{\rho_s}{\rho_l} )^{1/2}
\gamma_{lv} ) / (\rho_s L^b)$
corresponding to $T_m (R_c) = 0$, $L(R_c) = 0$,
below which the solid wire is not stable.

{\bf (ii) MD Simulation:} For our simulation we have chosen lead, and a
(110) orientation for the wires. Pb was a natural choice as the melting
properties of its surfaces and clusters have been extensively
investigated both experimentally and theoretically. Interestingly,
Pb(111), Pb(100) and Pb(110) exhibit different melting properties, that
is, respectively non-melting, incomplete melting and surface
melting \cite{bak1,bak2,bak3}. On small Pb particles, the formation of
a liquid skin as a premelting effect has been observed by microscopy
experiments \cite{french}. We have used a ``glue'' many-body potential,
already tested in previous surface \cite{tomagnini,goranka,toh} and
cluster \cite{lim} studies. In particular, the different melting behavior
of the main surfaces of Pb was found, in excellent agreement with
experimental results. Because the model potential is untested, and
therefore unreliable at really low coordination numbers, we restrict our
study to $R \ge 10\,\rm\AA$. By choosing a (110) wire axis,
(111), (100) and (110) orientations are simultaneously present on the
cylindrical surface of the wire. The different melting behavior of these
facets can make the details of premelting phenomena particularly
interesting.
Moreover, Pb(110) is the system on which the STM experiments
discussed above \cite{kuipers,kuipers1} have been done,
and one could conjecture that the neck might retain the same
(110) orientation of the sample surface.

We have performed several MD runs at different temperatures and for
different wire radii. All samples are made up by 36 (110) layers,
repeated along the wire axis $z$ by periodic boundary conditions.
The initial configurations are prepared by including all atoms
in an fcc lattice whose distance from the axis is smaller than an
assigned radius. At $T=0~K$, well defined (100),(110), and (111)
facets are present (Fig.~1). This roughly cylindrical geometry is
expected to be relatively close to the equilibrium situation, since
the surface energy anisotropy of Pb is known to be rather
small \cite{lim}. For the system sizes investigated, this procedure
typically results in octagonal cross sections. We have studied nine
samples, with a total number of atoms $N=$ 1134, 1206, 1314, 1710,
1854, 2286, 2574, 3294 and 4230, with radii between
$R= 13\,\rm\AA$ and $R= 25\,\rm\AA$.

All MD runs were performed at constant temperature and constant wire
length, where the box length along $z$ is adjusted according to the bulk
thermal expansion curve. While it would also seem possible to operate
at constant axial stress and varying length, it is not feasible to achieve
stability under such conditions when the wire becomes liquid.
The time step is $\Delta t=7.32\times 10^{-15}$~sec.
The initial samples with atoms in ideal fcc positions, were first relaxed
by  simple quenching. Then, an annealing cycle at room temperature,
followed by a quench to 0K, is done to check the stability of the original
structure.  In most cases, due to creation of defects during
the annealing, the final energy was slightly higher than that of the
previous relaxed structure. The absence of any major structural changes
indicates that the original structure was optimal or nearly optimal.
Each system was then heated by MD runs of 20000 time steps (147~ps)
with a temperature step of 50K up to 300K and of 25K above 300K.
In proximity of $T_m$ we further reduced the temperature step to
5K and used longer equilibration runs of 40000 time steps (293~ps)
at each temperature. We monitored several quantities, such as the
internal energy $E$, and the diffusivity as a function of $T$.

The caloric curves $E(T)$ for all the wires investigated are shown
in Fig.~\ref{caloric}. The dashed vertical line indicates the bulk melting
temperature of our model determined from bulk MD runs at zero pressure,
$T_m^b =$618K (to be compared with 600.7K of real Pb). In each curve
one can identify four regions, labeled as A, B, C, D
in Fig.~\ref{caloric}. In region A the wire is solid, and the slope
corresponds to the Dulong-Petit specific heat. Visual examination of
the samples confirms that diffusion is non--existent, or very low in
this region. In region B, $E(T)$ exhibits an upward curvature,
where the specific heat gradually increases by a small fraction of the
total heat of melting. This increase is associated with loss of solid
rigidity, and onset of diffusion at the wire surface.  For example,
the wire cross-section in Fig.~\ref{snap1}c, taken at $T=570$K for
a wire with $R= 22.5\,\rm\AA$, clearly shows that melting starts first
at the wire surface, while the inner region is still ordered.
The surface diffusivity is anisotropic and appears to be lower
in correspondence with (111) facets. This is reasonable in view
of the known non--melting behaviour of Pb(111)\cite{bak3,goranka}.
The lateral view (Fig.~\ref{snap1}d) also shows the presence of (111)
crystalline patches on the wire surface. Presumably, the wire size is
too small to exhibit complete crystallinity of (111) facets, which
should be expected in the limit of macroscopic sizes. (100) patches
are also sometimes observed.
(110) faces appear to be the first to melt with increasing $T$,
followed by (100) and then (111).
Another interesting feature showed by Fig.~\ref{snap1}c is
that (110) facets---and, to a lesser extent, (100) faces---have
shrunken in size with respect to the low-temperature shape,
and are considerably rounded.  A rounded profile is
characteristic of a roughened surface, where steps can
proliferate freely.  This observation is therefore qualitatively
consistent with the experimental observation of a roughening transition
of Pb(110) around 415K \cite{yang}, and is further corroborated by
surface simulations results \cite{toh}. Region C consists of an abrupt
nearly vertical jump corresponding to the melting transition of the
whole wire, and its temperature and magnitude are discussed below.
In region D the wire is completely liquid, and $E(T)$ grows
again linearly.

It is interesting to note that the internal energy jump (region C),
corresponding to the latent heat, remains sharp and discontinuous:
the wire suddenly melts when its melting temperature is reached,
regardless of the quantity of solid at the surface which is already
melted. This sharp jump yields a rather precise value of $T_m$
for all the sizes investigated. There is a clear decrease of $T_m$ with
decreasing wire radius (Fig.~\ref{caloric}). To verify the predicted
approximately factor of two of between Eqs.~(\ref{sctm}) and (\ref{cctm})
we compare the behavior of wires with that of Pb clusters investigated
by Lim {\em et al.} \cite{lim} using the same force model.
Figure~\ref{tm1r} reports $T_m$ as a function of $1/R$
for  wires (solid circles) and clusters (open squares).
The dashed lines correspond to the simple thermodynamical models of
Eqs.~(\ref{sctm}) and (\ref{cctm}),
using $\gamma_{sv} = 544\, \rm mJ/m^2$, $\gamma_{lv} = 476\, \rm mJ/m^2$,
$\rho_s = 0.0321\, \rm atom/\AA^3$, $\rho_l = 0.031\, \rm atom/\AA^3$ and
$L^b = 0.0494\, \rm eV/atom$.
For both geometries, the $1/R$ behavior is followed with fairly good
accuracy, as well as the approximate ratio 2 between the slopes.

The energy jump on melting corresponds to the latent heat
$L(R)$.  As a result of the premelting effects discussed
above, this jump is slightly smaller than the latent heat
that would be obtained if the system were entirely crystalline
just below $T_m$. In order to verify the prediction of Eq.~(\ref{ccl}),
where premelting effect are not considered, we have constructed a
modified latent heat $L^*(R)$, defined as the difference between the
energy of the liquid and the energy of a hypothetical complete solid
at $T_m$, obtained by ignoring surface melting, and extrapolating
linearly $E(T)$ from its low-temperature behavior.
$L^*(R)$ normalized by bulk latent heat, $L_b = 0.056\, \rm eV/atom$
determined from our model,
is plotted in Fig.\ \ref{lath} as a function of $1/R$
(solid circles), together with the data relative to spherical
clusters from ref.\ \cite{lim} (open squares) and
with Eq.~(\ref{ccl}) (dashed line).
For both geometries the $1/R$ behavior is again approximately
true, although not quite as accurate as for $T_m$. However,
deviations from Eq.~(\ref{ccl}) are more noticeable
for clusters than for wires, and for all the smaller sizes.
A faster decrease of $L$ compared with $T_m$ with
decreasing size has been also observed in previous Au clusters
simulations \cite{andreoni}.
These deviations can be tentatively attributed to the
reduced entropy of atoms at the liquid metal surface
compared with those in the bulk liquid.

Our results on wire melting may be of relevance to the neck between
an STM tip and a metal surface.
While a thick crystalline neck can still be slowly pulled out of a surface
via tip--induced diffusion, as explained in
Refs.~\cite{kuipers}~and~\cite{kuipers1}, it is possible, as indicated by
MD simulations \cite{tomagnini} that at least a part of the neck could be
extremely thin, $R \sim 5 \AA$.
Our extrapolated melting temperature for a Pb wire so very thin, is
lower than room temperature.
Hence, that neck portion would be completely liquid above 300K,
allowing matter to flow freely along its axis even under fast traction.
The possible existance of these ultra--thin necks might be checked
by looking for, e.g. transversally quantized electronic levels and related
jumps in the IV characteristics, and their liquid nature could be of
importance for friction problems.

In conclusion, thin wires, like small clusters, melt at a temperature
$T_m(R)$ which is lower than the bulk melting temperature, and is found
to be approximately a decreasing linear function of inverse radius.
The latent heat follows a similar law, but drops slightly faster with
decreasing size. Wire melting is preceded by surface melting effects of
its outer skin, similar to those exhibited by flat surfaces and by
spherical clusters. Possible relevance to the nature of thin tip--surface
necks has been pointed out.

\begin{figure}
\caption{Snapshot views of the MD sample with $N=3294$ and
$R\simeq 22.5\,\rm\AA$. a) and b) is the initial sample at 0K.
a)~is the top view of the wire cross section. Wire axis (z--direction)
is perpendicular to this cross section.
b)~is the side view from an angle such that all different laterals are
seen. c) and d) is corresponding views at a temperature just below the
melting temperature. The grayness of each atom is proportional to its
square displacement during a MD run. White indicates a non-diffusing,
crystalline behavior.
\label{snap1} }
\end{figure}
\begin{figure}
\caption{Internal energy as a function of temperature is shown for
different wire samples. Numbers labeling the curves are the number of atoms
in MD cell. Dashed vertical line indicates the bulk melting temperature for
lead determined from MD. Discontinuity in the curves shows the melting
transition.
\label{caloric} }
\end{figure}
\begin{figure}
\caption{Melting temperature of different clusters as a function of inverse
radius. Solid circles refer to wires, open squares to spherical clusters,
the dashed lines are the predictions of the simple phenomenological
theory (see text).
\label{tm1r} }
\end{figure}
\begin{figure}
\caption{Latent heat of melting for different clusters as a function of
inverse radius. Solid circles refer to wires, open squares to spherical
clusters, the dashed lines are the predictions of the simple
phenomenological theory (see text).
\label{lath} }
\end{figure}

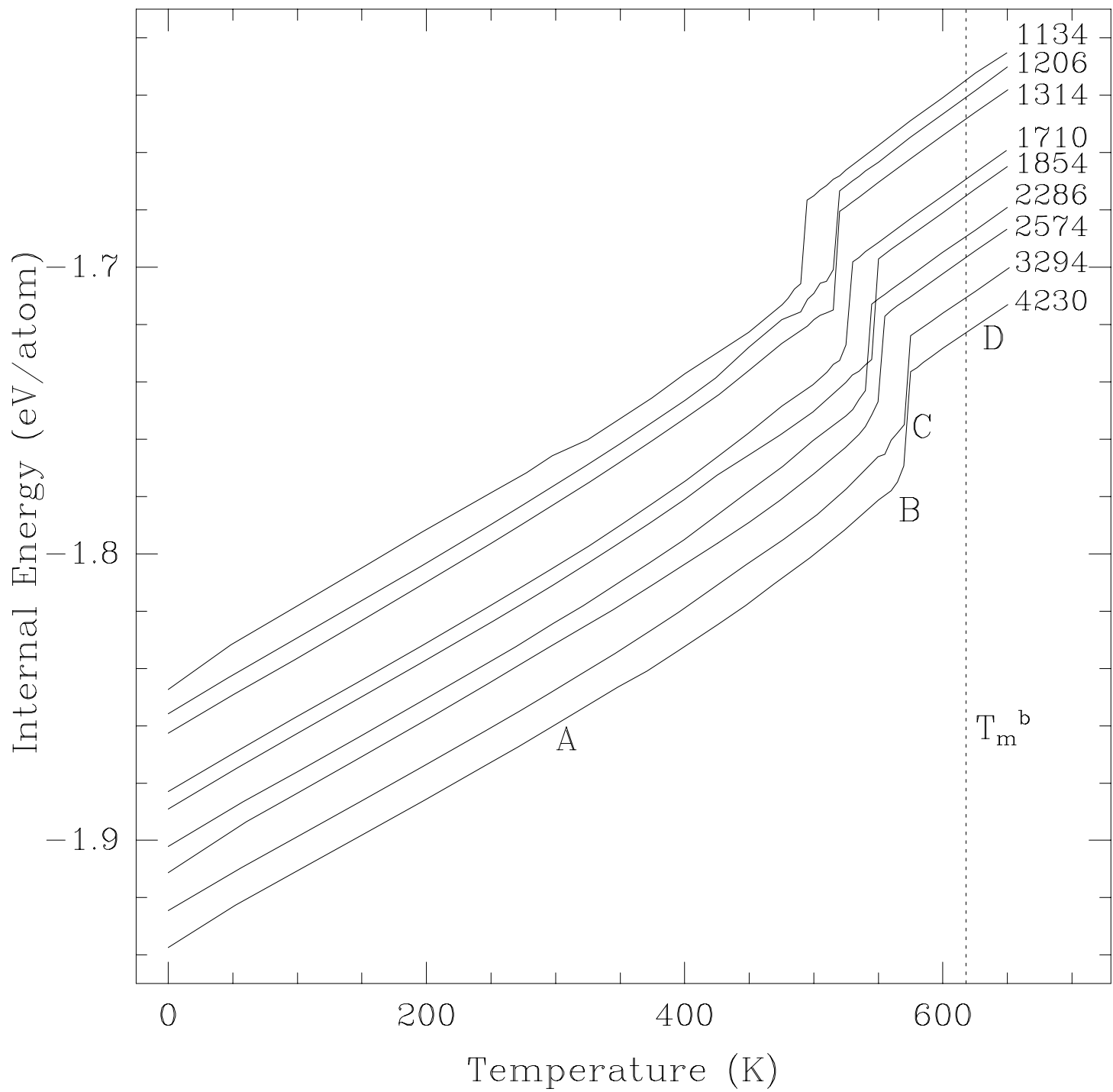
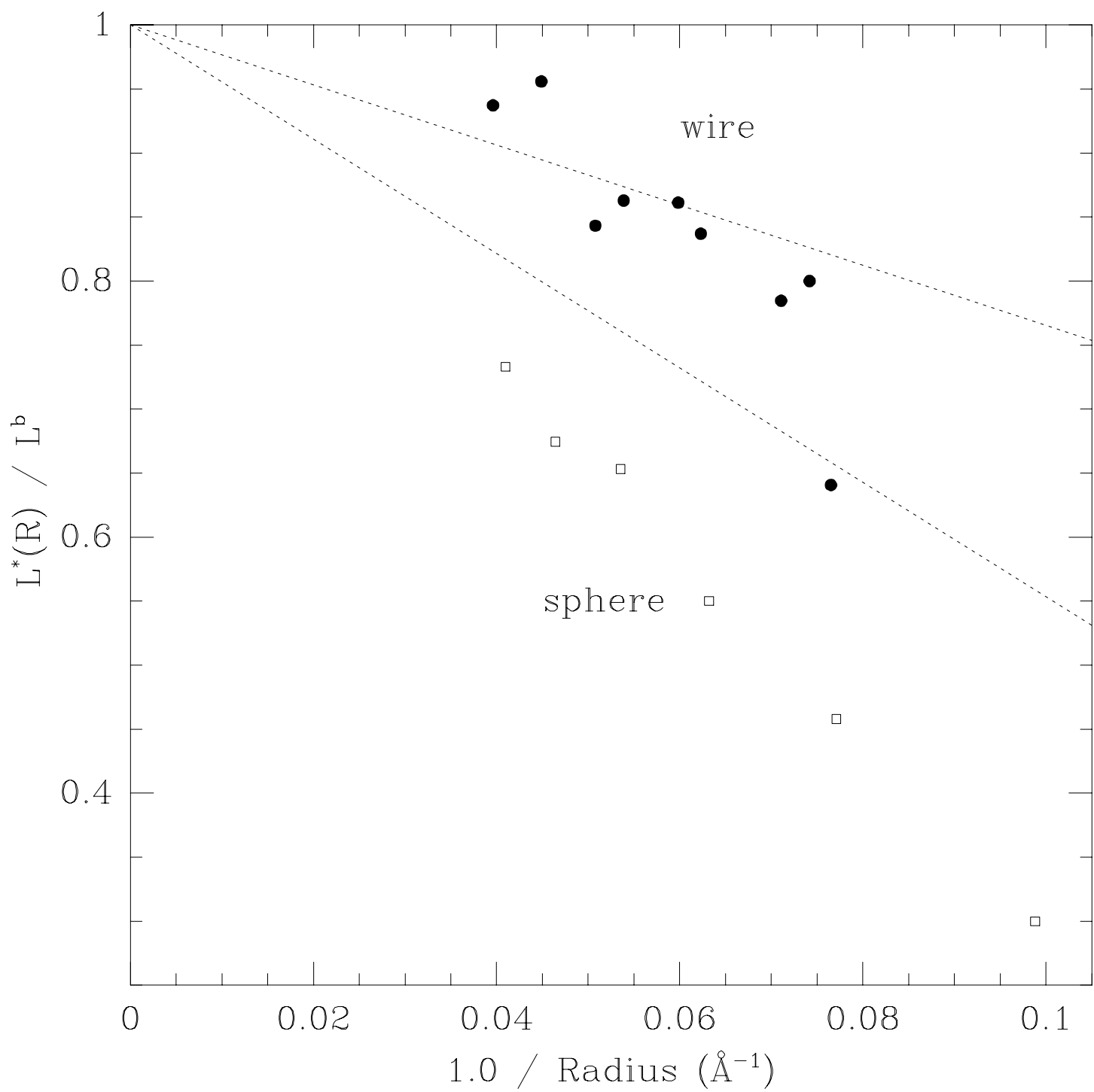
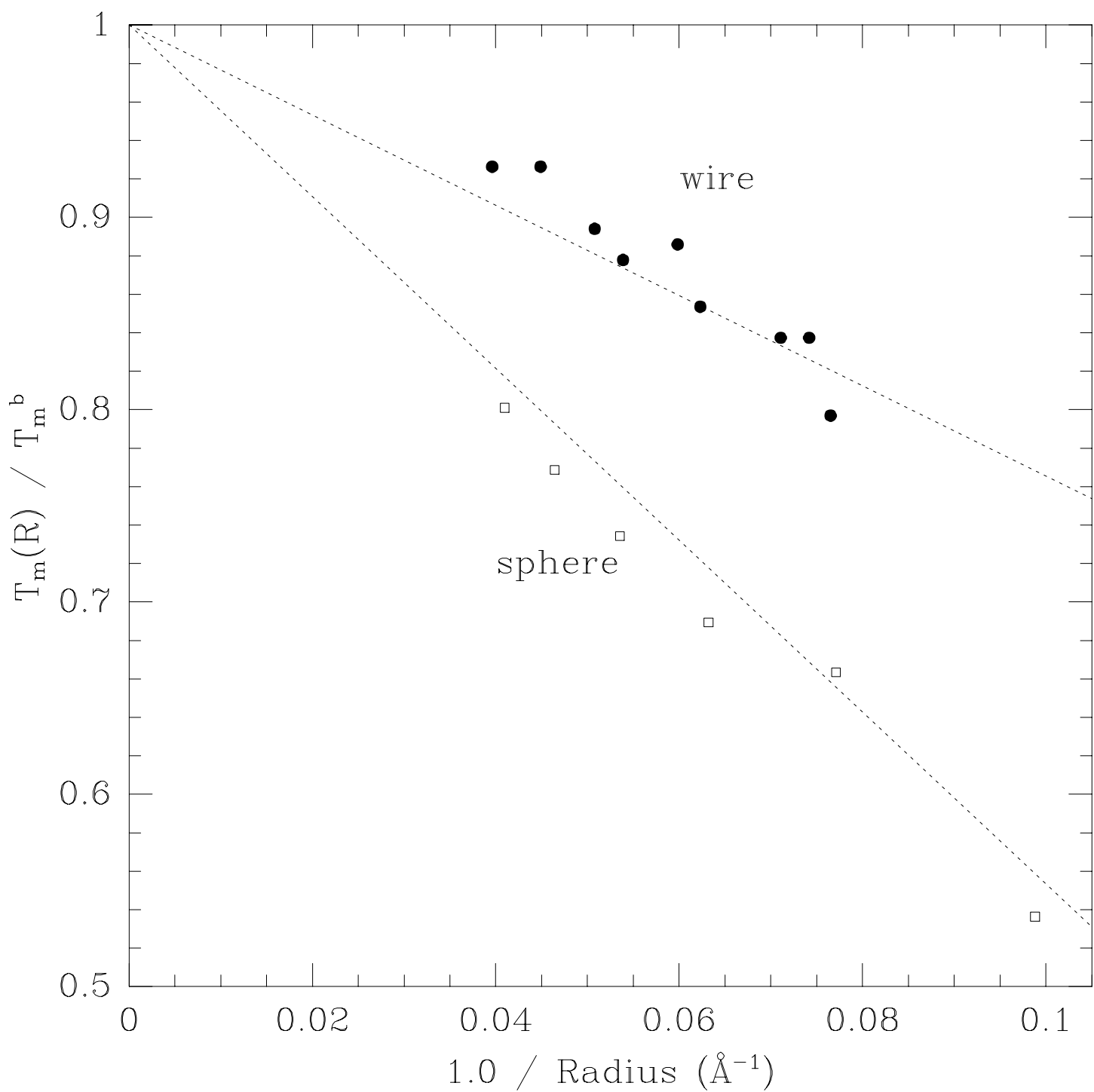
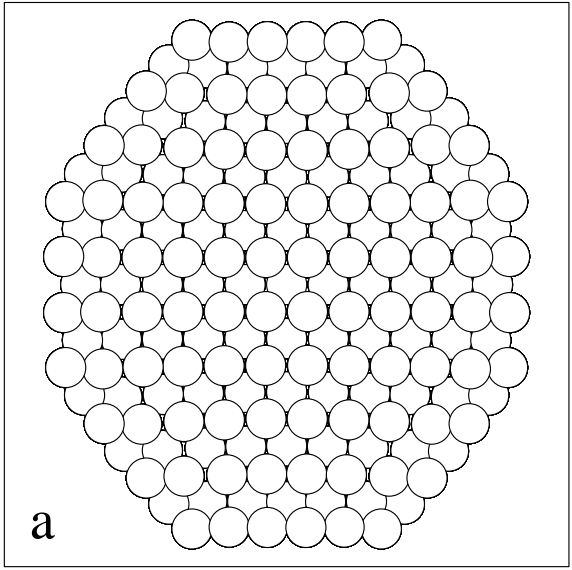
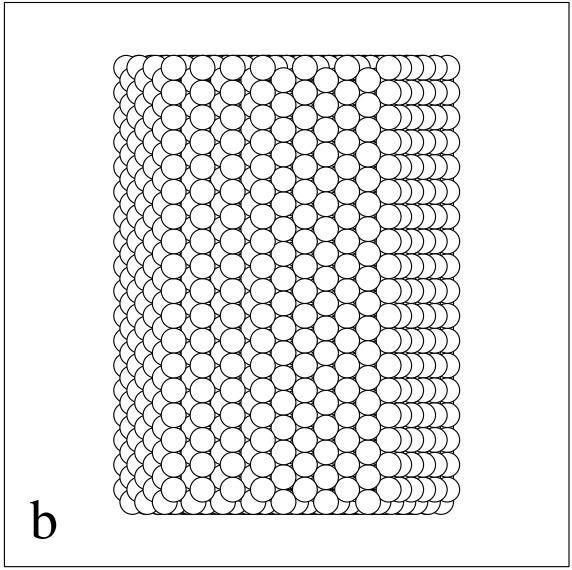
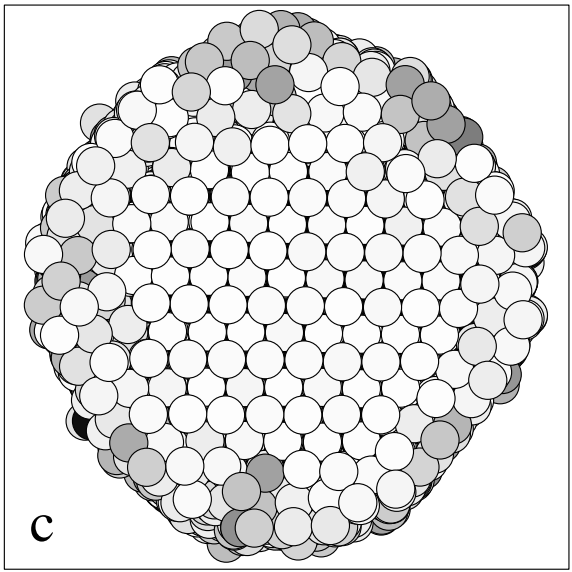
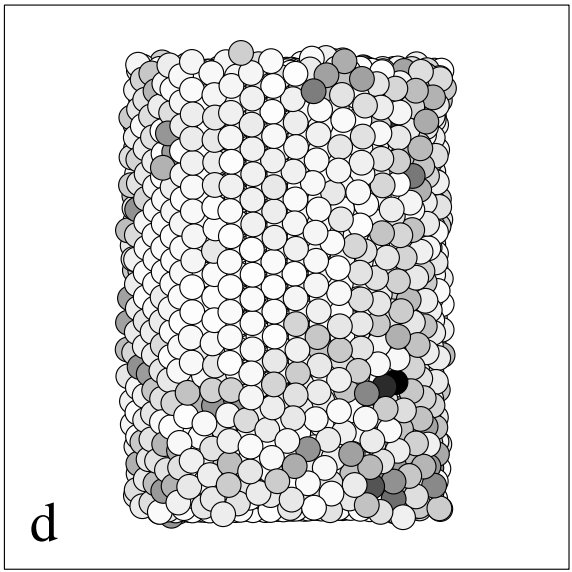
\end{document}